\documentclass[lettersize,journal]{IEEEtran}
\usepackage{amsmath,amsfonts,amssymb}
\usepackage{algorithmic}
\usepackage{algorithm}
\usepackage{array}
\usepackage{lipsum}
\usepackage{textcomp}
\usepackage{stfloats}
\usepackage{bm}
\usepackage{upgreek}
\usepackage{caption}
\usepackage{subcaption}
\usepackage{url}
\usepackage{verbatim}
\usepackage{graphicx}
\usepackage{cite}
\begin{document}

\title{Deep Learning-based Data-aided Activity Detection with Extraction Network in Grant-free Sparse Code Multiple Access Systems}

\author{{Minsig Han, Ameha T. Abebe, \IEEEmembership{Member, IEEE} and Chung G. Kang, \IEEEmembership{Senior Member, IEEE}}
\thanks{M. Han and C. Kang are with the School of Electrical Engineering, Korea University,
Seoul 136-701, South Korea (e-mail: als4585@korea.ac.kr; ccgkang@korea.ac.kr).
Ameha T. Abebe is with Samsung Research, Samsung Electronics, Seoul 06762, South Korea (e-mail: amehat.abebe@samsung.com).}}

\markboth{IEEE Wireless Communications Letters}
{Submitted paper}

\maketitle

\begin{abstract}
This letter proposes a deep learning-based data-aided active user detection network (D-AUDN) for grant-free sparse code multiple access (SCMA) systems that leverages both SCMA codebook and Zadoff-Chu preamble for activity detection. Due to disparate data and preamble distribution as well as codebook collision, existing D-AUDNs experience performance degradation when multiple preambles are associated with each codebook. To address this, a user activity extraction network (UAEN) is integrated within the D-AUDN to extract a-priori activity information from the codebook, improving activity detection of the associated preambles. Additionally, efficient SCMA codebook design and Zadoff-Chu preamble association are considered to further enhance performance.
\end{abstract}

\begin{IEEEkeywords}
Grant-free random access, sparse code multiple access, active user detection, deep learning, extraction network
\end{IEEEkeywords}

\section{Introduction}

\IEEEPARstart{G}{r}ant-free random access (GF-RA) is useful for massive machine type communication (mMTC) in IoT devices, allowing for sporadic uplink traffic. Grant-free non-orthogonal multiple access (GF-NOMA) uses code-domain non-orthogonal multiple access (CD-NOMA) to spread user data with non-orthogonal signatures for better overloading performance \cite{GFNOMASurvey, CBSCMA, PGTA}. Active users transmit a preamble and spread data using a randomly selected signature without scheduling, and receivers estimate each user's channel and detect transmitted data using the associated signature as long as active user preambles are detected \cite{CBSCMA}.

While it is possible to associate a unique preamble to an individual signature, i.e., one-to-one association, when there are many active users, a sufficient number of preambles is required to reduce a preamble collision rate. Due to the limited number of signatures as compared to the number of preambles, however, each signature may be associated with multiple preambles. This many-to-one association can lead to unavoidable performance degradation caused by signature collisions even when different preambles are selected \cite{CBSCMA,PGTA}. 
However, if the channel of the active user is accurately estimated and sufficiently separated, data can still be detected through the user-specific channel, implying that accurately estimating the channel of active users is crucial for data detection in GF-NOMA \cite{PGTA}.
On the other hand, accurate channel estimation (CE) heavily relies on the performance of the active user detection (AUD) procedure \cite{CSSurvey, PAUDN1, PAUDN2, DAUDN1, DAUDN2, DAUDN3}, which conventionally uses compressive sensing (CS) algorithms to exploit sparsity in user activity. However, these methods' performance deteriorates significantly with higher user activity rates \cite{CSSurvey}. Recent deep learning (DL)-based AUD schemes have been proposed to overcome this challenge, outperforming CS-based approaches while maintaining lower complexity \cite{PAUDN1, PAUDN2, DAUDN1, DAUDN2, DAUDN3}.


DL-based AUD networks (AUDNs) can be categorized as preamble-based (P-AUDN) \cite{PAUDN1, PAUDN2} or data-aided (D-AUDN) \cite{DAUDN1, DAUDN2, DAUDN3}. D-AUDN improves upon P-AUDN by using both preamble and data for activity detection. However, current implementations of D-AUDN only consider one-to-one preamble-signature association scenarios, and there is no existing implementation for many-to-one association scenarios. Using a single deep neural network (DNN) architecture for both preamble and signature activity detection can result in reduced training efficiency of D-AUDN \cite{DAUDN1, DAUDN2}. An alternative approach is to use a separate DNN structure for preamble and signature activity detection, which has been shown to improve training efficiency \cite{DAUDN3}. However, signature collision may result in a significant deterioration of D-AUDN's performance.

This letter proposes a novel D-AUDN structure for many-to-one preamble-signature association scenarios in GF-RA, using SCMA as a representative CD-NOMA scheme. SCMA's optimized low-density codebook (CB) allows for low-complexity multi-user detection through a message passing algorithm (MPA) \cite{SCMA1,SCMA2, SCMA3}. Inspired by the extraction network in \cite{Pilotless} that derives channel information from $M$-ary data, we propose a user activity extraction network (UAEN) that extracts user activity information from the spread and superposed data, improving activity detection of associated preambles in the long short-term memory (LSTM)-based AUDN. 
Our current design focuses on the receiver side for a fading channel, unlike the joint preamble and UAEN design in \cite{Globecom, APCC} which is subject to additive white Gaussian noise channel.
Additionally, we propose efficient SCMA CB structures and preamble association methods to enhance performance. Simulation results show that our proposed approach outperforms state-of-the-art AUDNs in GF-SCMA systems in terms of active users, signal-to-noise ratio (SNR), and activity detection error rate (ADER).

\begin{figure}[ht]
    \centering
    \includegraphics[width=3.5in,height=1.9in]{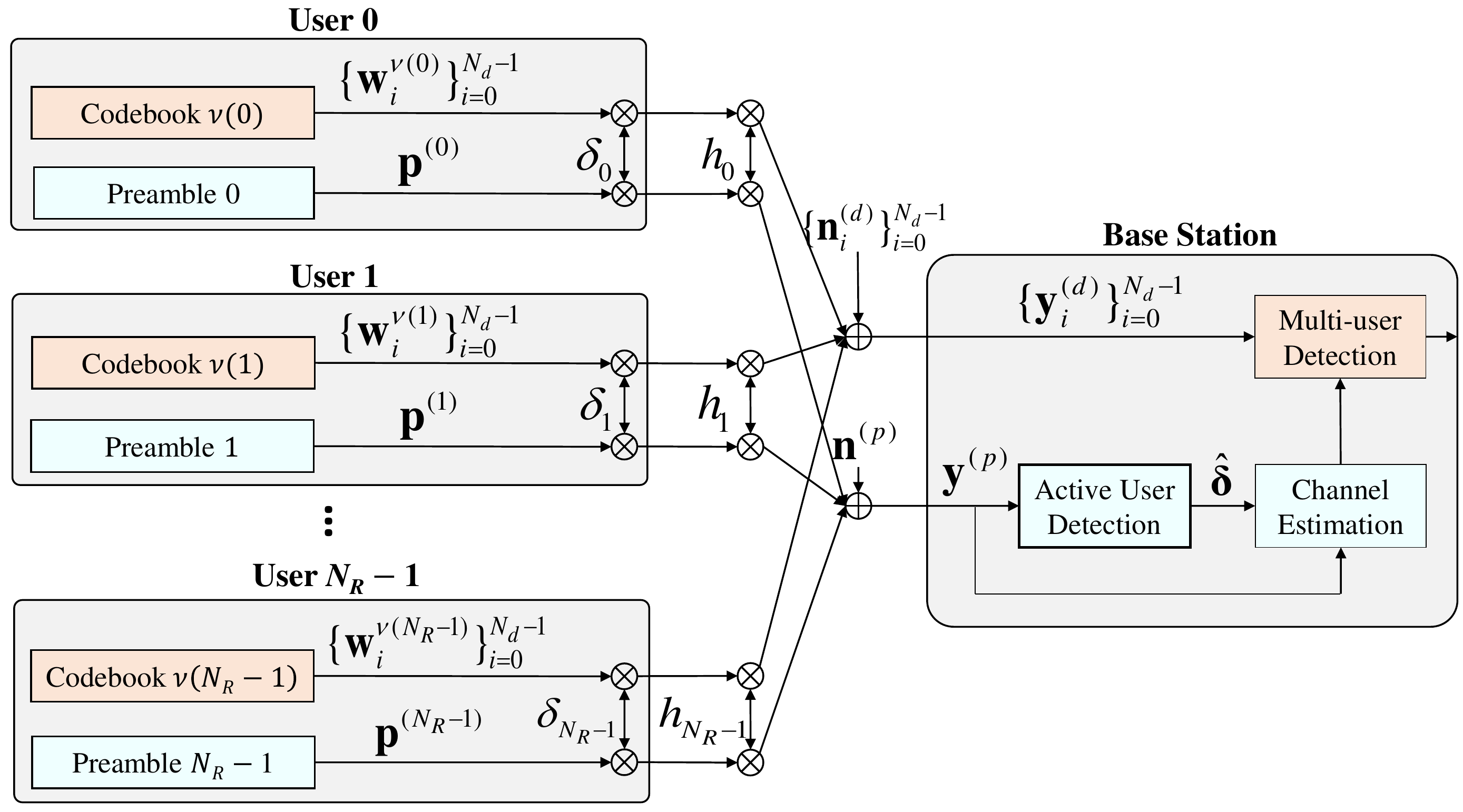}
    \caption{Transceiver model for GF-SCMA system}
    \label{fig_1}
\end{figure}

\section{System Model}
\subsection{Grant-free SCMA}
A total of ${{N}_{u}}$ users who have the potential to transmit, identified by their indices $n\in \{0,1,\ldots ,{{N}_{u}}-1\}$, are restricted to a specific time slot, which is shared subject to collision among them. Ultimately, only a subset of ${{N}_{u}}$ users will become active and transmit a preamble before sending a data burst to the time-synchronized base station. Assume that the activity of each user follows a Bernoulli distribution. More specifically, let ${{\delta }_{n}}\in \{0,1\}$ denote activity indicator of user $n$, i.e., ${{\delta }_{n}}$ = 1 indicating that user $n$ is active with a probability of ${{p}_{n}}$, and ${{\delta }_{n}}$ = 0 with a probability of ($1-{{p}_{n}}$). The activity indicators for all users constitute a random activity vector, represented as $\bm{\updelta }={{[{{\delta }_{0}},{{\delta }_{1}},\ldots ,{{\delta }_{{{N}_{u}}-1}}]}^{T}}$. 

For SCMA-based grant-free access with $J$ different CBs, let $\mathcal{C}{{}_{j}}$ denote the $j$-th $M$-ary CB, which is represented as ${{\mathcal{C}}_{j}}=\{\mathbf{c}_{0}^{(j)},\mathbf{c}_{1}^{(j)},\ldots ,\mathbf{c}_{M-1}^{(j)}\}$, $j=1,2,\cdots ,J$ where $\mathbf{c}_{m}^{(j)}\in {{\mathbb{C}}^{K}}$ denotes its $m$-th $K$-dimensional codeword (CW), e.g., employing $K$ subcarriers. Note that each CW is a sparse vector with $N$ non-zero elements ($N<K$). The $j$-th CB has an average power of ${{P}_{j}}=\left( {}^{1}/{}_{M} \right)\cdot \sum\nolimits_{m=0}^{M-1}{\left\| \mathbf{c}_{m}^{(j)} \right\|}_{2}^{2}$ and the total power of all CBs is $P=\sum\nolimits_{j=0}^{J-1}{{{P}_{j}}}$. In this context, there are two design options for SCMA CBs: the power-balanced CB (PB-CB), which distributes power evenly across all CBs \cite{SCMA1, SCMA2}, and the power imbalanced CB (PI-CB), which distributes power unevenly among CBs \cite{SCMA3}. 

Meanwhile, we consider non-orthogonal Zadoff-Chu (ZC) sequence-based preambles. A ZC sequence with length ${{N}_{ZC}}$, assuming ${{N}_{ZC}}$ is a prime number, and a root $u=\{1,2,\ldots ,{{N}_{ZC}}-1\}$ is given as 
\begin{equation}
\small
z{}_{u}(k)=\exp \left( {}^{-i\pi uk(k+1)}/{}_{{{N}_{ZC}}} \right)\,,k\text{=1,2,}\ldots \text{,}{{N}_{\text{ZC}}}    
\end{equation}
From a single root, ${{N}_{ZC}}$ orthogonal preambles can be generated through non-zero cyclic shift up to ${{N}_{ZC}}$. When more than ${{N}_{ZC}}$ preambles are required to maintain an acceptable level of preamble collision, ones from the different roots can be generated with their cross-correlation of $1/\sqrt{{{N}_{ZC}}}$. Accordingly, a total of ${{N}_{ZC}}({{N}_{ZC}}-1)$ non-orthogonal ZC preambles can be generated with a prime length ${{N}_{ZC}}$.

Assume that each CB is associated $L$ ZC preamble sequences. Therefore, we are considering a total of ${{N}_{R}}=J\cdot L$ contention transmission units (CTUs) \cite{CBSCMA}. In mMTC, the number of users is much larger than the number of available preambles, ${{N}_{R}}\ll {{N}_{u}}$. Therefore, in order to minimize the probability of preamble collisions, it is advantageous to associate as many preambles as possible \cite{CBSCMA}.

\subsection{Virtual Transceiver Model for DL-based AUD Design}
Due to a massive number of users, it is challenging to optimize AUD for all users while taking potential collisions into account. As in the existing DL-based AUD designs, therefore, we adopt a similar design approach optimized for all possible ${{N}_{R}}$ preambles in the absence of collisions \cite{PAUDN1, PAUDN2, DAUDN1, DAUDN2, DAUDN3}. Toward this end, we consider a virtual transceiver model as shown in Fig. 1, which deals with ${{N}_{R}}$ simultaneously users, indexed by $n\in \{0,1,\ldots ,{{N}_{R}}-1\}$. Let ${{\mathbf{p}}_{n}}\in {{\mathbb{C}}^{{{N}_{ZC}}}}$ represent a ZC preamble assigned to user $n$. Assigning a unique non-orthogonal ZC preamble to each user, it intends to train DL-based AUD subject to the highly overloaded situation.

As $J<{{N}_{R}}$, CBs must be reused for many-to-one preamble-CB association, suffering from multi-user interference. We assume that ${{N}_{R}}$ users employ a means of round-robin CB selection, i.e., user $n$ to select a CB with index $\nu (n)=\bmod (n,J)$. Then, let ${{\mathcal{C}}_{\nu (n)}}$ denote a CB for user $n$ where $\nu (n)=\bmod (n,J)$. To average out a CB collision rate, homogeneous user activity is assumed by ${{p}_{0}}={{p}_{1}}=\ldots ={{p}_{{{N}_{R}}-1}}=\bar{p}$ \cite{Globecom}. Then the expected number of active users within a time slot, denoted by ${{N}_{a}}$, is given as ${{N}_{a}}={{N}_{u}}\bar{p}$.

A bit sequence of active user is divided into ${{N}_{d}}$ blocks, each with ${{\log }_{2}}M$ bits. Then, each block is encoded with its own preassigned CB. Let $\mathbf{b}_{i}^{(n)}\in {{\mathbb{B}}^{{{\log }_{2}}M}}$ denote the $i$-th block of user $n$, which is encoded with a CW, denoted as $\mathbf{w}_{i}^{\nu (n)}\in {{\mathcal{C}}_{\nu (n)}}$. Assuming that CTU transmissions are experiencing a flat fading channel, let ${{h}_{n}}$ denote a channel coefficient for user $n$, thus constituting a channel vector for all users, represented as $\mathbf{h}={{[{{h}_{0}},{{h}_{1}},\ldots ,{{h}_{{{N}_{R}}-1}}]}^{T}}$. Note that a flat fading channel model is commonly accepted for the mMTC users with no movement \cite{CSSurvey,PAUDN1, PAUDN2, DAUDN1, DAUDN2, DAUDN3}. Consequently, the received signals, ${{\mathbf{y}}^{(p)}}$ and $\mathbf{y}_{i}^{(d)}$, for the preamble and the $i$-th CW superposed over the given resources are expressed respectively as
\begin{equation}
\small
{{\mathbf{y}}^{(p)}}=\sum\nolimits_{n=0}^{{{N}_{R}}-1}{{{\delta }_{n}}{{h}_{n}}{{\mathbf{p}}^{(n)}}}+{{\mathbf{n}}^{(p)}},
\end{equation}
and
\begin{equation}
\small
\mathbf{y}_{i}^{(d)}=\sum\nolimits_{n=0}^{N-1}{{{\delta }_{n}}{{h}_{n}}\mathbf{w}_{i}^{\nu (n)}}+\mathbf{n}_{i}^{(d)},\text{  }i\text{=}0,\ldots ,{{N}_{d}}-1
\end{equation}
where ${{\mathbf{n}}^{(p)}}\sim \mathcal{C}\mathcal{N}(0,{{\sigma }^{2}}{{\mathbf{I}}_{{{K}^{(p)}}}})$ and $\mathbf{n}_{i}^{(d)}\sim \mathcal{C}\mathcal{N}(0,{{\sigma }^{2}}{{\mathbf{I}}_{K}})$ are the additive white Gaussian noise (AWGN) vector  with $M\times M$ identity matrix ${{\mathbf{I}}_{M}}$. In Fig. 1, conventional CS-based AUD can be performed only using the superposed preamble ${{\mathbf{y}}^{(p)}}$ to detect an activity vector, represented as $\bm{\hat{\updelta }}={{[{{\hat{\delta }}_{0}},{{\hat{\delta }}_{1}},\ldots ,{{\hat{\delta }}_{{{N}_{R}}-1}}]}^{T}}$. After the AUD process, CE and multi-user detection (MUD) for active users can be performed using a linear minimum mean square estimator and MPA, respectively \cite{CBSCMA}.

The objective of the DL-based AUDs is to optimize an AUD network (AUDN), denoted as non-linear mapping $g(\cdot )$, in terms of its trainable parameters \cite{PAUDN1, PAUDN2, DAUDN1, DAUDN2, DAUDN3}. More specifically, the AUDN must be designed to minimize the ADER, which can be formulated as the following optimization problem:
\begin{equation}
\small
{{g}^{*}}=\underset{g}{\mathop{\arg \min }}\,{{\left\| \bm{\updelta }-g(\cdot ;{{\bm{\uptheta }}^{(g)}}) \right\|}_{0}}
\end{equation}
where ${{\bm{\uptheta }}^{(g)}}$ is a vector of weight and bias in AUDN and ${{\left\| \cdot  \right\|}_{0}}$ denotes L-0 norm. Depending on whether it is preamble-based AUDN (P-AUDN) \cite{PAUDN1, PAUDN2} or data-aided (D-AUDN) \cite{DAUDN1, DAUDN2, DAUDN3}, DNN is represented as $g(\cdot ;{{\bm{\uptheta }}^{(g)}})={{g}_{p}}({{\mathbf{y}}^{(p)}};{{\bm{\uptheta }}^{({{g}_{p}})}})$ or $g(\cdot ;{{\bm{\uptheta }}^{(g)}})={{g}_{d}}({{\mathbf{y}}^{(p)}},\{\mathbf{y}_{i}^{(d)}\}_{i=0}^{{{N}_{d}}-1};{{\bm{\uptheta }}^{({{g}_{d}})}})$, respectively, where ${{\bm{\uptheta }}^{({{g}_{k}})}}$ is a vector of weight and bias of ${{g}_{k}}(\cdot )$. Both ${{g}_{p}}({{\mathbf{y}}^{(p)}};{{\bm{\uptheta }}^{({{g}_{p}})}})$ and ${{g}_{d}}({{\mathbf{y}}^{(p)}},\{\mathbf{y}_{i}^{(d)}\}_{i=0}^{{{N}_{d}}-1};{{\bm{\uptheta }}^{({{g}_{d}})}})$ are trained to detect user activity $\bm{\hat{\updelta }}$, i.e., $\bm{\hat{\updelta }}={{g}_{p}}({{\mathbf{y}}^{(p)}};{{\bm{\uptheta }}^{({{g}_{p}})}})$ or $\bm{\hat{\updelta }}={{g}_{d}}({{\mathbf{y}}^{(p)}},\{\mathbf{y}_{i}^{(d)}\}_{i=0}^{{{N}_{d}}-1};{{\bm{\uptheta }}^{({{g}_{d}})}})$. In GF-SCMA, since ${{\mathbf{y}}^{(p)}}$ and $\{\mathbf{y}_{i}^{(d)}\}_{i=0}^{{{N}_{d}}-1}$ over the same resources are received from the same active users, their joint distributions can be exploited to improve the AUDN performance.

\section{Proposed Design of Data-aided AUD Network (D-AUDN)}

\subsection{Overall Structure}
Fig. 2(a) shows the proposed architecture for D-AUDN with a user activity extraction network (UAEN) that extracts activity information jointly from $\{\mathbf{y}_{i}^{(d)}\}_{i=0}^{{{N}_{d}}-1}$ and ${{\mathbf{y}}^{(p)}}$. In order to exploit both ${{\mathbf{y}}^{(p)}}$ and $\{\mathbf{y}_{i}^{(d)}\}_{i=0}^{{{N}_{d}}-1}$ as input for the AUDN, it needs to be transformed into a double-length real vector by concatenating its real and imaginary parts. The UAEN is designed to generate a priori probabilities of user activity for individual users, denoted as a vector $\bm{\upalpha }=[{{\alpha }_{0}},{{\alpha }_{1}},\cdots ,{{\alpha }_{{{N}_{R}}-1}}]$, where ${{\alpha }_{n}}$ represents a probability that user $n$ is active given $\{\mathbf{y}_{i}^{(d)}\}_{i=0}^{{{N}_{d}}-1}$. More specifically, the UAEN is represented by the following nonlinear mapping $h(\{\mathbf{y}_{i}^{(d)}\}_{i=0}^{{{N}_{d}}-1};{{\bm{\uptheta }}^{(h)}})=\bm{\upalpha }$ where ${{\bm{\uptheta }}^{(h)}}$ is a vector of weight and bias in the network. The detailed design approach is discussed in the sequel.

As $\bm{\upalpha }$ is exploited as a priori information for our proposed D-AUDN, the proposed AUDN can be represented by the following nonlinear mapping ${{g}_{\alpha }}(\bm{\upalpha },{{\mathbf{y}}^{(p)}};{{\bm{\uptheta }}^{({{g}_{\alpha }})}})=\bm{\hat{\upeta }}$ where ${{\bm{\uptheta }}^{({{g}_{\alpha }})}}$ is a vector of weight and bias in the proposed D-AUDN and $\bm{\hat{\upeta }}=[{{\hat{\eta }}_{0}},{{\hat{\eta }}_{1}},\cdots ,{{\hat{\eta }}_{{{N}_{R}}-1}}]$. The user activity $\bm{\hat{\updelta }}$ is finally determined by a threshold criterion, i.e., user $n$ is declared active if ${{\hat{\eta }}_{n}}>\gamma $, where $\gamma $ denotes an activity threshold \cite{PAUDN1,PAUDN2,DAUDN1,DAUDN2,DAUDN3}. A serially concatenated vector of $\bm{\upalpha }$ and ${{\mathbf{y}}^{(p)}}$ is taken as an input to the D-AUDN ${{g}_{\alpha }}(\cdot )$. As $\bm{\upalpha }$ is user activity information extracted from $\{\mathbf{y}_{i}^{(d)}\}_{i=0}^{{{N}_{d}}-1}$, the proposed D-AUDN is expected to improve the performance through activity information jointly extracted from both data and preamble.

\begin{figure}[htp]
\captionsetup[subfigure]{justification=centering}
    \begin{subfigure}[b]{\linewidth}
        \centering
        \includegraphics[width=3.5in]{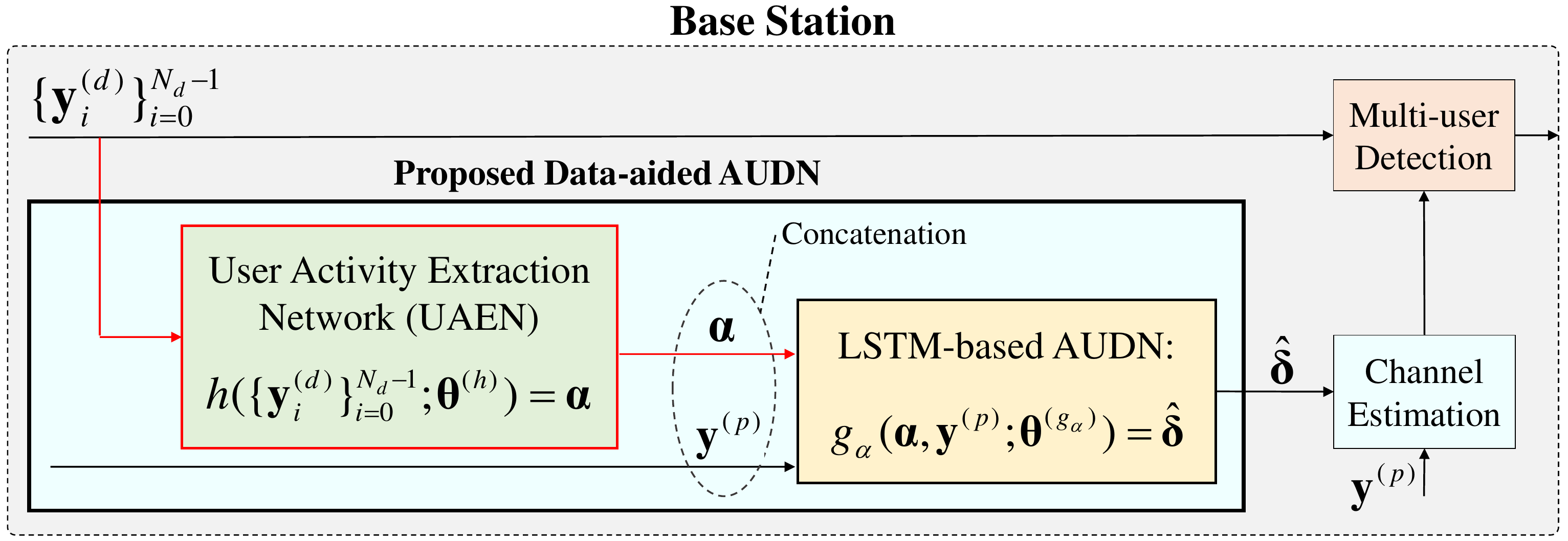}
        \caption{Overall Receiver Structure}
        \label{fig:a}
    \end{subfigure} %
    \hfill
    \begin{subfigure}[b]{\linewidth}    
        \includegraphics[width=3.5in]{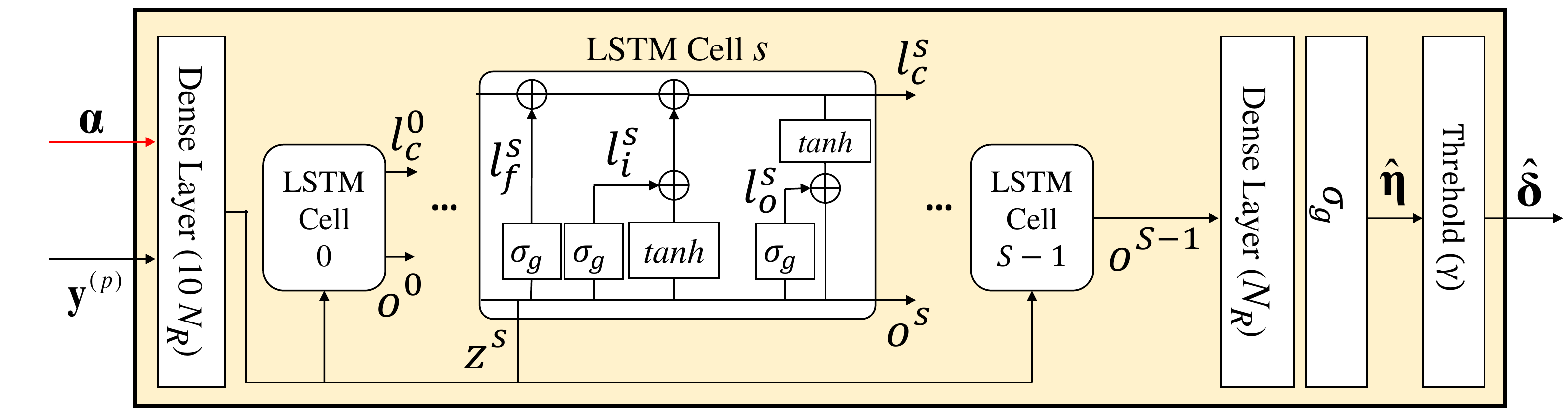}
        \caption{Detailed Structure of LSTM-based AUDN}
        \label{fig:b}    
    \end{subfigure} 
    \caption{Structure of Proposed D-AUDN}
\end{figure}

As shown in Fig. 2(b), ${{g}_{\alpha }}(\cdot )$ utilizes the LSTM-based AUDN which was adopted in \cite{DAUDN2}. It uses the input and forget gates of the LSTM network to select and reject active and inactive users from NOMA measurements, which results in better convergence and generalization performance compared to residual network (ResNet)-based AUDNs \cite{PAUDN2, DAUDN1}. This is particularly true when input sparsity varies dynamically over time. The LSTM-based AUDN structure is detailed in the sequel.

The $s$-th LSTM cell consists of cell-state $l_{c}^{s}$ and three gates, i.e., forget gate $l_{f}^{s}$, input gate $l_{i}^{s}$, and output gate $l_{o}^{s}$. Let $\mathbf{z}_{{}}^{s}$ and $\mathbf{o}_{{}}^{s}$ denote the input and output of the $s$-th cell. These gating operations are defined as $l_{x}^{s}={{\sigma }_{g}}({{\mathbf{W}}_{x}}{{\mathbf{z}}^{s}}+{{\mathbf{U}}_{x}}{{\mathbf{o}}^{s-1}}+{{\mathbf{b}}_{x}})$, $x=f,i$ and $o$, where ${{\sigma }_{g}}(x)=[1/(1+{{e}^{-x}})]$ denotes sigmoid function, ${{\mathbf{W}}_{x }}\in {{\mathbb{R}}^{10{{N}_{R}}\times 10{{N}_{R}}}}$ and ${{\mathbf{U}}_{x}}\in {{\mathbb{R}}^{10{{N}_{R}}\times 10{{N}_{R}}}}$ are the weights associated with ${{\mathbf{z}}^{s}}$ and $\mathbf{o}_{{}}^{s-1}$, respectively, while ${{\mathbf{b}}_{x}}\in {{\mathbb{R}}^{10{{N}_{R}}\times 1}}$ is the bias. Then, a cell-state $l_{c}^{s}$ of the $s$-th cell is given by $l_{c}^{s}=l_{f}^{s}\circ l_{c}^{s-1}+l_{i}^{s}\circ \bar{l}_{c}^{s}$ where $\bar{l}_{c}^{s}=\tanh ({{\mathbf{W}}_{c}}{{\mathbf{z}}^{s}}+{{\mathbf{U}}_{c}}{{\mathbf{o}}^{s-1}}+{{\mathbf{b}}_{c}})$ while $\circ $ denotes the Hadamard product. Meanwhile, the output of the $s$-th LSTM cell is given by ${{\mathbf{o}}^{s}}=l_{o}^{s}\circ \tanh (l_{c}^{s})$ \cite{DAUDN2}.

Ideally, it is possible to jointly train the AUDN and UAEN only by BCE loss function in \cite{PAUDN1, PAUDN2, DAUDN1, DAUDN2}, which is given as
\begin{equation}
\small
\mathcal{L}(\bm{\updelta },\bm{\hat{\upeta }})=-\sum\nolimits_{n=0}^{N-1}{\{{{\delta }_{n}}\log ({{{\hat{\eta }}}_{n}})+(1-{{\delta }_{n}})\log (1-{{{\hat{\eta }}}_{n}})\}}.
\end{equation}
However, the deep location of the UAEN component within AUDN causes gradient vanishing during backpropagation, leading to poor convergence performance when using equation (5). To address this issue, a pre-training scheme and a specific UAEN structure are proposed in the following subsections.

\subsection{Design of User Activity Extraction Network (UAEN)}
Processing $\{\mathbf{y}_{i}^{(d)}\}_{i=0}^{{{N}_{d}}-1}$ jointly with UAEN requires expanding the size of UAEN, which may lead to the vanishing gradient problem and high computational complexity. To address these issues, one possible solution is to reduce the number of data samples by splitting the processing of $\{\mathbf{y}_{i}^{(d)}\}_{i=0}^{{{N}_{d}}-1}$ into independent smaller networks. The outputs of these smaller networks can then be combined later in an ensemble network. This approach can help manage the computational complexity and mitigate the vanishing gradient problem.

Fig. 3 shows the proposed UAEN employs two 1D-convolutional layers to produce an intermediate activity vector. The first convolutional layer is made of $N_{\text{kernel}}^{(1)}$ kernels with a size of $2K$. Let ${{\bm{{\alpha }''}}_{i}}\in {{\mathbb{R}}^{N_{\text{kernel}}^{(1)}}}$ represent its output vector for the first kernels from $\mathbf{y}_{i}^{(d)}$. As there tends to be too many trainable parameters for the dense network of combining the intermediate activity vectors, we consider another convolutional layer that can reduce the number of trainable parameters. The second convolutional layer to reduce $\{{{\bm{{\alpha }''}}_{i}}\}_{i=0}^{{{N}_{d}}-1}$ is made of $N_{\text{kernel}}^{(2)}$ kernels with a size of $N_{\text{kernel}}^{(1)}$, where $N_{\text{kernel}}^{(1)} > N_{\text{kernel}}^{(2)}$. Let ${{\bm{{\alpha }'}}_{i}}\in {{\mathbb{R}}^{N_{\text{kernel}}^{(2)}}}$ represent its output vector for ${{\bm{{\alpha }''}}_{i}}\in {{\mathbb{R}}^{N_{\text{kernel}}^{(1)}}}$, forming the intermediate activity vector. After constructing $\{{{\bm{{\alpha }'}}_{i}}\}_{i=0}^{{{N}_{d}}-1}$, a single dense layer is applied to the UAEN with ${{N}_{R}}$ nodes to ensemble them. Finally, another dense layer with sigmoid function ${{\sigma }_{g}}(\cdot )$ is adopted in the output to produce the activity vector $\bm{\upalpha }\in {{\mathbb{R}}^{{{N}_{R}}}}$. Note that batch normalization and ReLU activation function are used after each layer for the improved convergence performance.

\begin{figure}
\centering
\includegraphics[width=3.5in,height=1.85in]{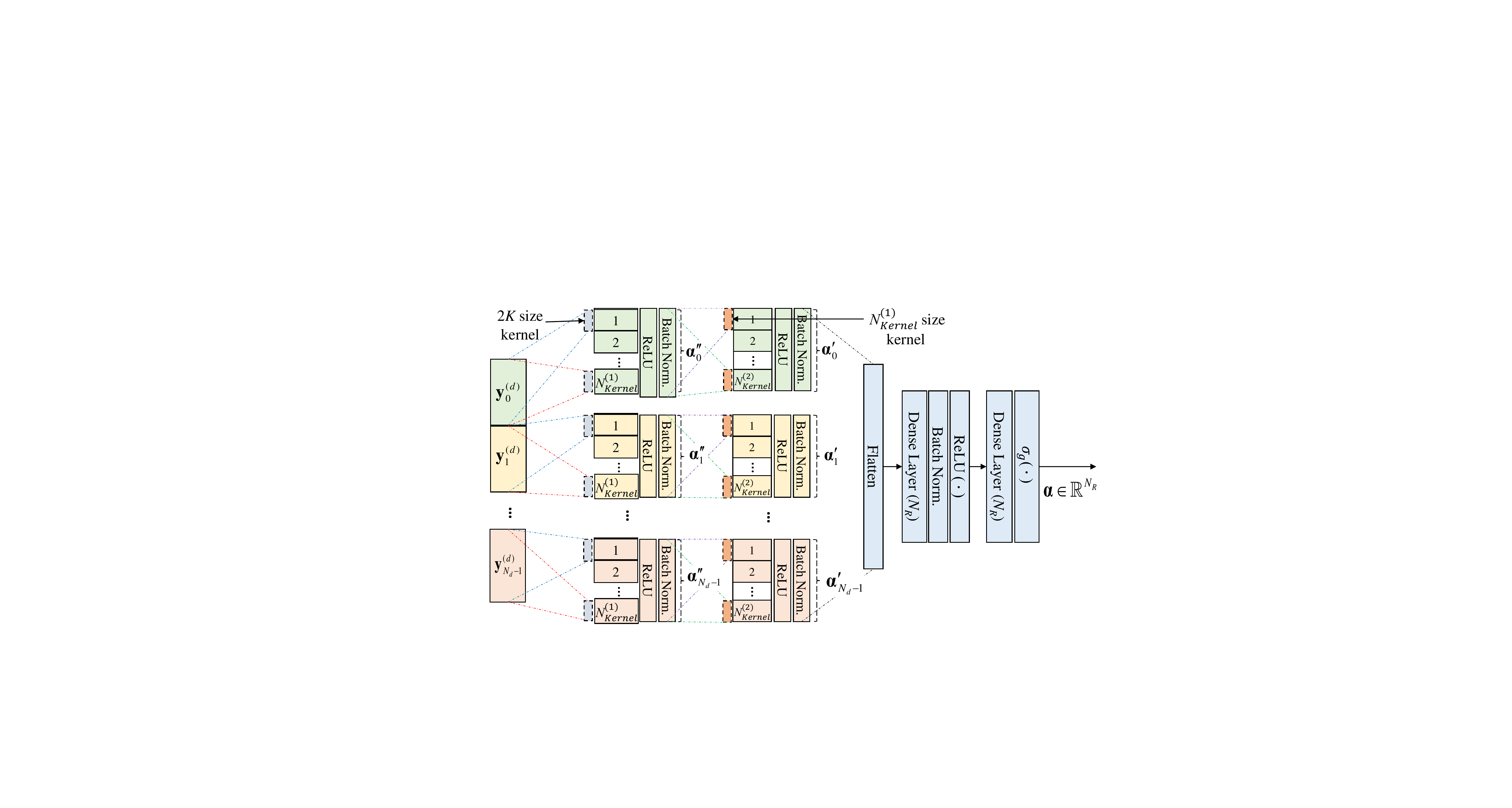}
\caption{Detailed Structure of UAEN}
\label{fig_1}
\end{figure}

\subsection{Progressive Training for proposed D-AUDN}
D-AUDN is deeper than existing DL-based AUDNs as it includes an additional network for extracting user-activity information, but may suffer from the vanishing gradient problem if only joint training with the loss function in (5) is used. To address this, a progressive training approach is employed where the UAEN component is pre-trained before joint UAEN and AUDN training.

A self-supervised loss function is employed for pre-training the UAEN in such a way that its output $\bm{\upalpha }$ becomes close to the activity vector $\bm{\updelta }$ as follows:
\begin{equation}
\small  
{{\mathcal{L}}_{PT}}(\bm{\updelta },\bm{\upalpha })=-\sum\nolimits_{n=0}^{N-1}{\{{{\delta }_{n}}\log ({{\alpha }_{n}})+(1-{{\delta }_{n}})\log (1-{{\alpha }_{n}})\}}.
\end{equation}
To comprehensively incorporate the AUDN, fine-tuning of the UAEN through joint training with equation (5) is necessary, as using equation (6) for pre-training alone does not consider the AUDN. This joint training incorporates $\bm{\upalpha }$ as the priori user activity information for AUDN to minimize ADER. A two-step training approach is employed that combines pre-training and joint training, as described in \cite{Globecom}.

\section{CTU Design for Proposed Data-aided AUDN}

\subsection{Optimizing CB Design for D-AUDN }
Detecting CB activity directly from superposed data in fading channels is a challenging task, especially for the original SCMA CB design \cite{SCMA1} that employs constellation rotation. This design is vulnerable to phase changes in the fading channel, leading to potential loss of CB activity information. Resource sparsity may differentiate CBs, but it becomes harder with more active users.

To overcome this challenge, we have explored the use of CBs with asymmetric constellations in the same resource, such as PB-CB \cite{SCMA2} and PI-CB \cite{SCMA3}. Among these, PI-CB offers two significant advantages over PB-CB for D-AUDN. Firstly, it provides a greater minimum Euclidean distance, aiding the D-AUDN in distinguishing CB activity. Secondly, the asymmetry of the constellation, induced by different average CB power levels, is useful for the UAEN in extracting CB activity in a fading channel. Asymmetric constellation points have also been shown to be useful in estimating time-varying fading channels without pilots in \cite{Pilotless}.

\subsection{Root-separated Zadoff-Chu Association}
To ensure an effective CB-preamble association for CTU design, both CB distance and preamble distance should be considered together to maximize the minimum distance between CTUs. The distance between preambles can be measured using cross-correlation, with higher cross-correlation indicating a closer distance \cite{CSSurvey}. As CBs with the same index have a closer distance, two CTUs with the same CB must reply on the preambles with a greater distance, i.e., associating the same CB with the preambles of the farther distance (lower cross-correlation), as discussed in \cite{APCC}. Toward this end, each CB is assigned to a distinct set of orthogonal ZC preambles generated from the same root sequence, with the different roots evenly distributed among all CBs. This root-separated (RS) ZC preamble association improves a distance between active CTUs belonging to the same CB.

\section{Simulation Results}
We consider a GF-SCMA system with $J=6$, $K=4$, and $N=2$. Two SCMA constellation designs, i.e., PB-CB in \cite{SCMA2} or PI-CB in \cite{SCMA3}, are considered for the data transmission. The number of CWs used for data transmission is set to ${{N}_{d}}=16$ \cite{Globecom}. Considering two different lengths of the ZC preambles, i.e., ${{N}_{ZC}}=7$ or ${{N}_{ZC}}=13$, orthogonal preambles from the six roots, i.e., $u=\{1,2,\ldots ,6\}$, are utilized to generate a total number of $6{{N}_{ZC}}$ non-orthogonal preambles. Those ZC preambles are associated with $J=6$ SCMA CBs, either randomly or using RS ZC association in Section IV-B.

We assume independent and identically distributed Rayleigh flat-fading channel with the SNR of 20dB for training data set. We assume that the number of active users, $N_a$, is uniformly distributed over $[1,6]$. Data samples are constructed by activity vectors $\bm{\updelta }$ for the given number of active users. We generate $5\cdot10^5$ activity vectors in the pre-training step, while $5\cdot10^6$ activity vectors are used in the joint training step. The learning rates of 0.001 and 0.0001 for pre-training and joint training, respectively, and the training epoch is set to 50. We employ an ADAM optimizer with a batch size of 1000 \cite{DAUDN2}. Since it is verified that DL-based AUDs have superior ADER performance and lower computational complexity compared to conventional CS-based AUDs, we will focus on how much performance gain can be achieved compared to DL-based AUD designs, i.e., P-AUDN in \cite{PAUDN2} and D-AUDN in \cite{DAUDN2}. The number of LSTM cells for AUDN and activity threshold are set to $S=10$ and $\gamma = 0.5$, respectively \cite{DAUDN2}. We selected sufficiently large hyperparameters for UAEN, i.e., $N_{\text{kernel}}^{(1)}=10{{N}_{R}}$ and $N_{\text{kernel}}^{(2)}=2{{N}_{R}}$, through hyperparameter tuning in Rayleigh fading channel. The number of trainable parameters in the UAEN are only $2.43\%$ and $4.67\%$ of those in AUDN for ${{N}_{R}}=42$ and ${{N}_{ZC}}=78$, respectively.

\begin{figure}
\captionsetup[subfigure]{justification=centering}

    \begin{subfigure}[b]{0.495\linewidth}
        \includegraphics[width=4.35cm,height=4.0cm]{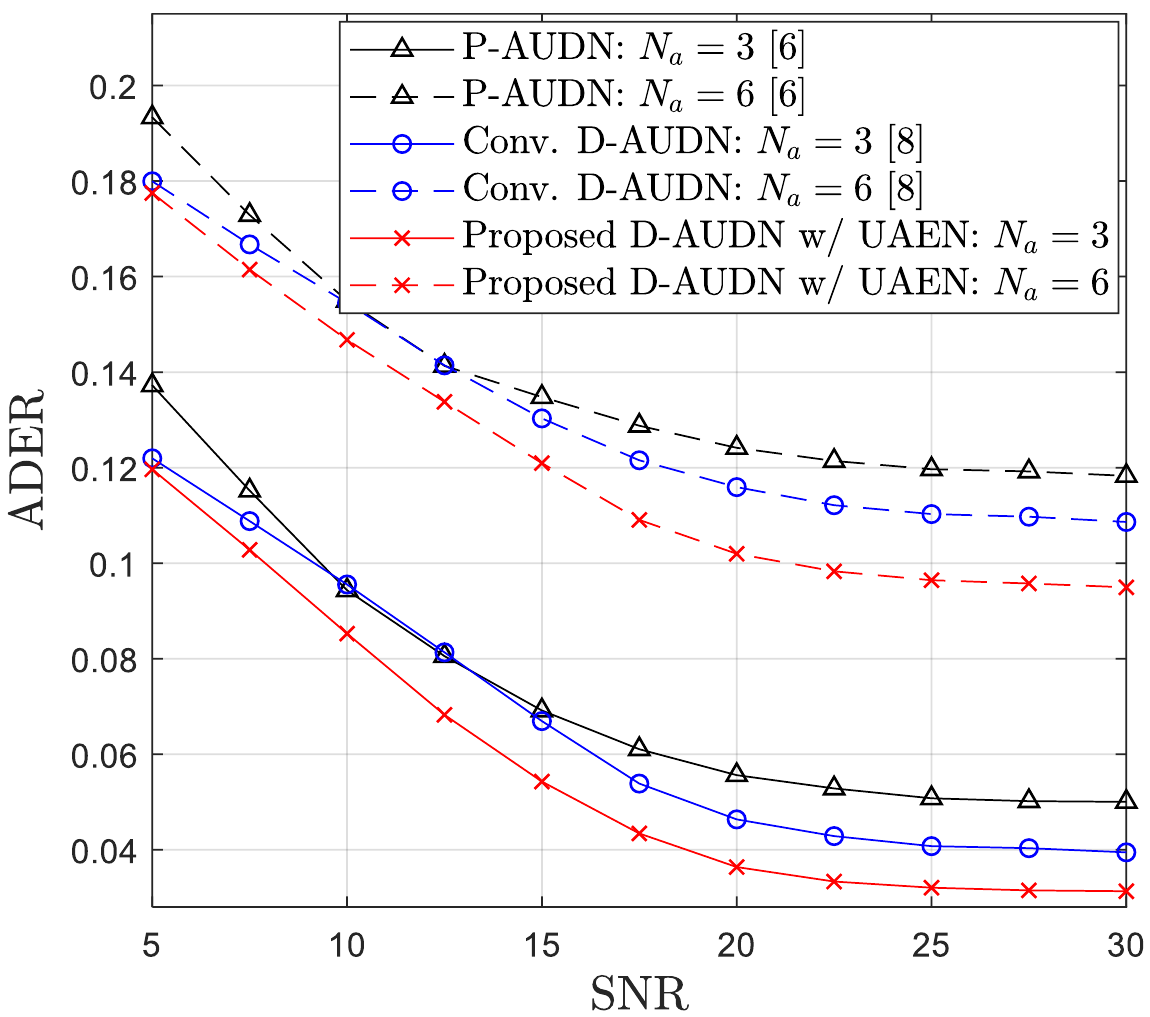}
        \caption{ADER vs. SNR}
        \label{fig:a}
    \end{subfigure} %
    \hfill
    \begin{subfigure}[b]{0.495\linewidth}    
        \includegraphics[width=4.35cm,height=4.0cm]{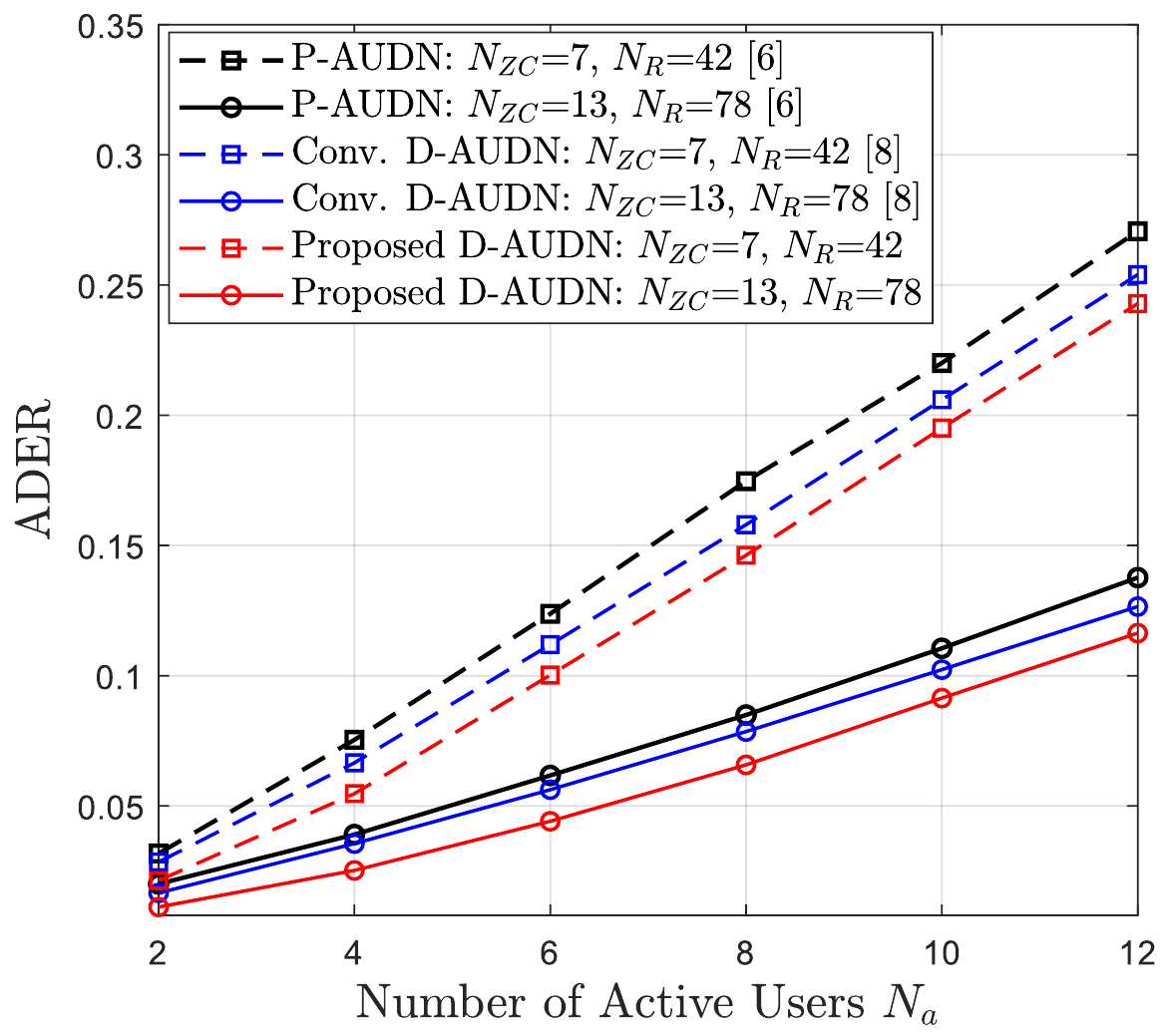}
        \caption{ADER vs. $N_a$}
        \label{fig:b}    
    \end{subfigure} 
    \caption{ADER performance comparison with existing AUDNs}
\end{figure}

Fig.’s 4(a) and 4(b) compare the ADER performance of the proposed AUDN with that of the existing DL-based AUDN in \cite{PAUDN2} and \cite{DAUDN2}, considering the PB-CB and random ZC association. In Fig. 4(a), ADER performance is presented as varying the SNRs in a flat-fading channel with ZC preamble length ${{N}_{ZC}}=7$. The proposed D-AUDN outperforms both P-AUDN and conventional D-AUDN in a wide range of SNR in both ${{N}_{a}}=3$ and ${{N}_{a}}=6$. Meanwhile, Fig. 4(b) shows the ADER as varying the number of active users at 20dB SNR. Note that the proposed D-AUDN is trained with $N_{a}=[1,6]$, while the ADER is evaluated with a wider range of $N_{a}=[2,12]$. It has been observed that the proposed D-AUDN can robustly support additional active users compared to P-AUDN at ADER of 0.05 when using short ZC sequences with a length of 7 and 13, respectively, while outperforming the conventional D-AUDN approach. This suggests that a robust data-aided gain can be achieved from the proposed D-AUDN structure.

In Fig. 5, we examined the impact of proposed CTU design on D-AUDN performance in ${{N}_{a}}=6$, comparing two different CB designs, PB-CB and PI-CB, respectively. In addition, we evaluate the efficiency of RS ZC preamble association proposed in Section IV-B by comparing it with random ZC preamble association. Specifically, in RS ZC preamble association, preambles generated from the 6 roots are separately associated with one of the six CBs. Fig. 5 indicates that using PI-CB results in around 2dB gain at ADER of 0.1 and 0.05, compared to PB-CB in ${{N}_{ZC}}=7$ and 13, respectively. Furthermore, utilizing RS ZC preamble association can provide up to 1dB and 0.5dB performance gain over the random ZC preamble association in ${{N}_{ZC}}=7$ and 13, respectively. Note that the shorter ZC preambles result in higher non-orthogonality, causing a smaller minimum CTU distance when randomly associated. As a result, the performance gain from RS ZC association is more significant with shorter ZC sequences.

\begin{figure}
\captionsetup[subfigure]{justification=centering}

    \begin{subfigure}[b]{0.495\linewidth}
        \includegraphics[width=4.35cm,height=4.0cm]{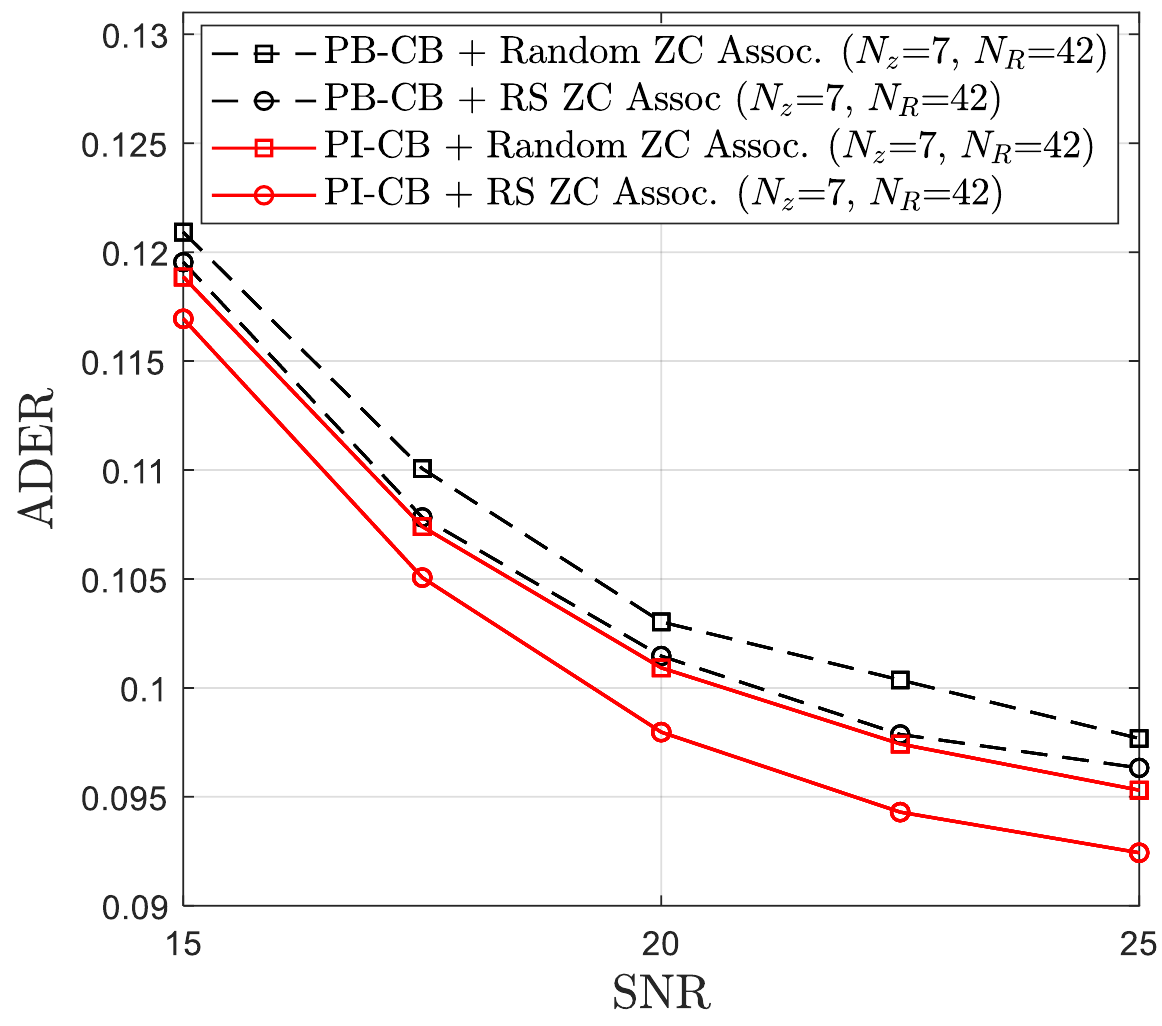}
        \caption{$N_{ZC}=7$}
        \label{fig:a}
    \end{subfigure} %
    \hfill
    \begin{subfigure}[b]{0.495\linewidth}    
        \includegraphics[width=4.35cm,height=4.0cm]{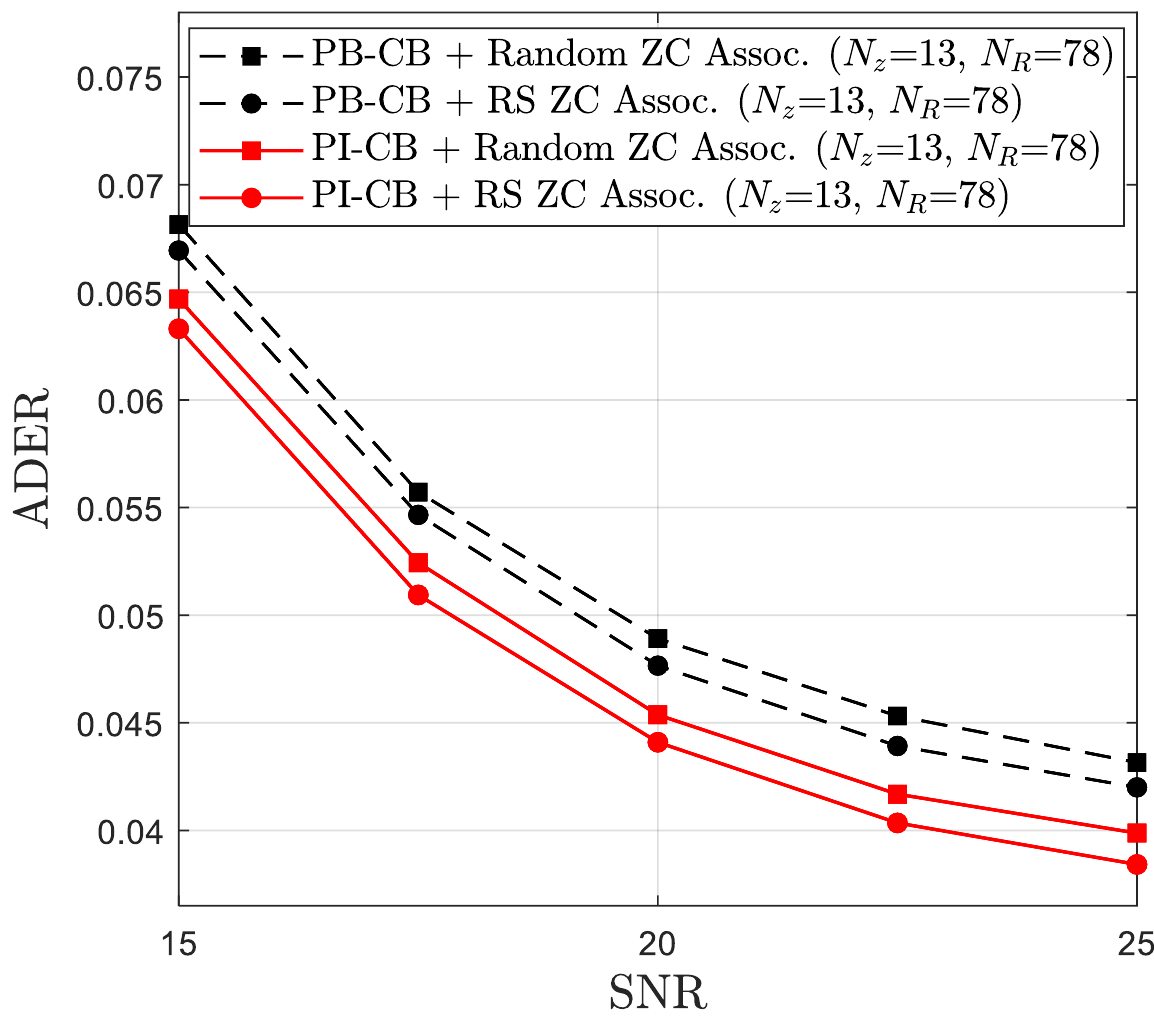}
        \caption{$N_{ZC}=13$}
        \label{fig:b}    
    \end{subfigure} 
    \caption{Impact of CTU Design for proposed D-AUDN: $N_a=6$}
\end{figure}

\section{Conclusions}
We proposed a DL-based approach for D-AUDN in GF-SCMA systems, utilizing a user activity extraction network that efficiently exploits SCMA CBs and Zadoff-Chu preamble for activity detection. Additionally, we used power-imbalanced SCMA CB and a root-separated ZC preamble association method to further improve D-AUDN performance. Our proposed structure can also be applied to other types of GF-NOMA systems requiring many-to-one association between preamble and signature. Future research could extend our proposed architecture to integrate with CE and MUD in highly overloaded mMTC systems, where challenges such as preamble non-orthogonality and signature collisions are present.

\bibliographystyle{ieeetr}
\bibliography{ref}

\end{document}